\documentclass[a4paper,11pt]{article}
\usepackage{graphicx}
\usepackage{fullpage}
\usepackage{jheppub} % for details on the use of the package, please
\usepackage{booktabs}
\usepackage{subcaption}
\usepackage{framed}
\usepackage{multirow}
\usepackage{array}
\usepackage{lscape}
\usepackage{rotating}
\usepackage{float}

\def\be{\begin{equation}}
\def\ee{\end{equation}}
\def\bea{\begin{array}}
\def\eea{\end{array}}
\def\beqa{\begin{eqnarray}}
\def\eeqa{\end{eqnarray}}
\def\beqas{\begin{eqnarray*}}
\def\eeqas{\end{eqnarray*}}

\def\bp{\begin{picture}}
\def\ep{\end{picture}}
\def\bc{\begin{center}}
\def\ec{\end{center}}
\def\bfig{\begin{figure}}
\def\efig{\end{figure}}

\def\bit{\begin{itemize}}
\def\eit{\end{itemize}}
\def\nn{\nonumber}
\def\f{\frac}

\def\[{\left[}
\def\]{\right]}
\def\({\left(}
\def\){\right)}

\def\..{\left.}
\def\.{\right.}
\def\tl{\tilde}
\def\ra{\rightarrow}
\def\la{\leftarrow}

\def\tm{\times}

\def\NPB#1,{{ Nucl.\ Phys.\ B }{\bf #1},}
\def\PLB#1,{{ Phys.\ Lett.\ B }{\bf #1},}
\def\EPJC#1,{{ Eur.\ Phys.\ Jour.\ C }{\bf #1},}
\def\PRD#1,{{ Phys.\ Rev.\ D }{\bf #1},}
\def\PRL#1,{{ Phys.\ Rev.\ Lett.\ }{\bf #1},}
\def\MPLA#1,{{Mod.\ Phys.\ Lett.\ A }{\bf #1},}

\def\da{\dagger}

\def\la{\lambda}

\def\al{\alpha}

\def\ep{\epsilon}

\title{750 GeV diphoton resonance, 125 GeV Higgs and muon $g-2$ anomaly in deflected anomaly mediation SUSY breaking scenarios}

\author[a,b]{Fei Wang,}
\author[c]{Lei Wu,}
\author[b,d]{Jin Min Yang,}
\author[b]{Mengchao Zhang}

\affiliation[a]{School of Physics, Zhengzhou University, Zhengzhou 450000, China}
\affiliation[b]{Key Laboratory of Theoretical Physics, Institute of Theoretical Physics, Academia Sinica, Beijing 100190, China}
\affiliation[c]{ARC Centre of Excellence for Particle Physics at the Terascale, School of Physics, The University of Sydney, NSW 2006, Australia}
\affiliation[d]{Department of Physics, Tohoku University, Sendai 980-8578, Japan}

\emailAdd{feiwang@zzu.edu.cn}
\emailAdd{leiwu@itp.ac.cn}
\emailAdd{jmyang@itp.ac.cn}
\emailAdd{mczhang@itp.ac.cn}

\abstract{
We propose to interpret the 750 GeV diphoton excess in deflected anomaly mediation supersymmetry breaking
scenarios,
which can naturally predict couplings between a singlet field and vector-like messengers.
The CP-even scalar component ($S$) of the singlet field can serve as the 750 GeV resonance.
The messenger scale, which is of order the gravitino scale, can be as light as $F_\phi \sim {\cal O}(10)$ TeV
when the messenger species $N_F$ and the deflection parameter $d$ are moderately large.
Such messengers can induce the large loop decay process $S \to \gamma\gamma$. Our results show that
such a scenario can successfully accommodate the 125 GeV Higgs boson, the 750 GeV diphoton excess and
the muon $g-2$ without conflicting with the LHC constraints.
We also comment on the possible explanations in the gauge
mediation supersymmetry breaking scenario.}

\begin{document}
\maketitle \indent
\newpage

\section{Introduction}
Very recently, the ATLAS and CMS collaborations have reported a resonance-like excess at 750 GeV in the diphoton invariant mass spectrum at the 13 TeV LHC \cite{750atlas,750cms}. Combined with the 8 TeV data, the production rate of the diphoton excess is given by \cite{ex-0}
\begin{equation}
\sigma^{750 GeV}_{\gamma\gamma} = (4.4\pm 1.1) ~\rm{fb}~.\label{excess}
\end{equation}
Although the local significance of this excess is only about $3\sigma$, many theoretical explanations for this excess have been proposed \cite{ex-1,ex-2,ex-3,ex-4}.

Among various extensions of the Standard Model (SM), the low energy supersymmetry (SUSY) is widely regarded as one of the most appealing candidates for new physics at the TeV scale. It can successfully overcome the gauge hierarchy problem encountered in the SM and also provide a compelling cold dark matter candidate. More intriguingly, the observed 125 GeV Higgs boson \cite{ATLAS:higgs,CMS:higgs} and the muon $g-2$ measurement \cite{g-2} can be naturally accommodated in some low energy SUSY models \cite{125-susy}. If SUSY is indeed the new physics beyond the SM, it should also explain the recently reported 750 GeV diphoton excess.

On the other hand, since no strong evidences of sparticles are found, the SUSY breaking scale has been pushed up to several TeV. This leads to a challenge for constructing feasible SUSY breaking
mechanisms. Among them, the deflected anomaly mediation SUSY breaking (AMSB) mechanism
\cite{deflect,deflect:RGE-invariance} is an elegant solution, which solves the tachyonic slepton
problem \cite{tachyonslepton} in the minimal AMSB \cite{AMSB} by introducing the messenger sector.
Besides, if the general messenger-matter interactions are introduced in the deflected AMSB scenario,
several other benefits can be obtained, such as the prediction of 125 GeV Higgs boson and the explanation of the muon $g-2$ anomaly \cite{fei-PLB}. In this work, we propose to interpret the 750 GeV diphoton excess in the deflected anomaly mediation SUSY breaking scenario, which contains a singlet superfield $\hat{S}$ and vector-like messengers. The CP-even scalar component $S$ of the singlet superfield can serve as the 750 GeV resonance. When the messenger species $N_F$ and the deflection parameter $d$ are moderately large, the messenger fields can be as light as ${\cal O}(10)$ TeV and can enhance the diphoton decay process $S \to \gamma\gamma$.

The paper is organized as follows. In Section II, we discuss the feasibility that the messenger scale can be as light as ${\cal O} (10)$ TeV in certain extensions of deflected AMSB scenario. In Section III, we perform numerical calculation and interpret the 750 GeV diphoton excess in our scenario. Finally, we draw our conclusions and comment on the explanation in gauge mediated SUSY breaking scenario \cite{GMSB}.

\section{Deflected anomaly mediation scenario}
In deflected AMSB scenario, vector-like messengers are introduced to deflect the  Renormalization Group Equation trajectory. The simplest possibility is given by \cite{deflect}
\beqa\label{superpotential}
W=\sum\limits_{i=1}^{N_F}\la_P X \tl{P}_i P_i~,
\eeqa
where $P_i,\tl{P}_i$ are messenger fields in terms of SU(5) fundamental (or antisymmetric {\bf 10}) representation
with following decomposition in term of $SU(3)_c\tm SU(2)_L\tm U(1)_Y$
  \beqa
  P_i({\bf 5})&=&{\bf (~1,  2)_{-1/2}\oplus (~3,~1)_{1/3}}~,
P_i({\bf 10})={\bf (~3,  2)_{-1/6}\oplus (~\bar{3},~1)_{2/3}\oplus (~1,~1)_{-1}} \\
  \tl{P}_i(\overline{\bf {5}})&=&{\bf (~1,  \bar{2})_{1/2}\oplus (~\bar{3},~1)_{-1/3}}~,
 \tl{P}_i(\overline{\bf 10})={\bf  (~\bar{3},  2)_{1/6}\oplus (~{3},~1)_{-2/3}\oplus (~1,~1)_{1}}
     \eeqa
After minimization of the SUSY version of Coleman-Weinberg potential, this theory gives a deflection parameter
\beqa
d\equiv\f{F_{\tl{X}}}{\tl{X}F_\phi}\approx -1,
\eeqa
with $\tl{X}=X\phi$. The purpose of the deflection is to solve the tachyonic slepton problem
in the minimal AMSB scenario. A numerical study indicates that non-tachyonic slepton masses require the messenger species to be larger than 4 for very heavy messengers with ${\bf 5}\oplus {\bf \bar{5}}$ representations (a very large number of messenger species may cause the gauge couplings to meet the Landau pole before the Planck scale).

On the other hand, if certain superpotential for $X$ is introduced,
the deflection parameter could be ${\cal O}(1)$ and takes either sign. In fact, the positively deflected AMSB scenario can be realized with the typical values of $X$ exponential \cite{okada} or with large couplings \cite{WWYZ}. An alternative way to evade the decoupling theorem in AMSB \cite{spinner} is to extend the anomaly mediation scenario by introducing the holomorphic Kahler potential. Such a holomorphic Kahler potential can naturally arise by integrating out heavy fields at tree-level. The simplest
feasible way to include a holomorphic Kahler potential \cite{weiner} is through the following interactions,
\beqa
\Delta L&=&\int d^4\theta \f{\phi^\da}{\phi} \(\sum\limits_{i}c_P^i P_i \tl{P}_i+c_S \hat{S}^2\)+\int d^2\theta W(\hat{S},P,\tl{P})+h.c.~,\nn\\
&=&-|F_\phi^2| (c_P^i  P_i \tl{P}_i +c_S \hat{S}^2)+ \int d^2\theta  F_\phi^\da \(c_P^i P_i \tl{P}_i+ c_S \hat{S}^2 \) +h.c.+\cdots~.
\eeqa

With $\phi=1+\theta^2 F_\phi$, we can see that the mass terms for the scalar component of $\hat{S}$ (denoted as $S$ )
will give tachyonic eigenvalues for $|2c_S|<1$. Such a tachyonic scalar can be stabilized
by the superpotential of $\hat{S}$ with its lowest component VEV $\langle S\rangle$ at the order of $F_\phi$. We choose the following superpotential with
the coupling between the singlet $\hat{S}$ and the messenger fields \cite{luty}
\beqa\label{potential}
W=\sum\limits_{i} \la_P^i \hat{S} \tl{P}_i P_i+ \f{\la_S}{3} \hat{S}^3~.
\eeqa
Here we neglect the possible UV divergent linear term of $\hat{S}$ \cite{luty2}.  Note that the coupling $S\tl{P}P$ in AMSB is different from the coupling $X\tl{P}P$ in GMSB. In GMSB, the singlet that couples to the messengers acquires F-term VEV from the hidden sector. While in AMSB type scenario, the SUSY breaking information is encoded in the compensator field $\phi=1+\theta^2 F_\phi$ and $S$ acts differently with respect to $X$.

Adding the superpotential term from the Kahler part to Eq.(\ref{potential}), we can obtain
 \beqa
 -F_{\hat{S}}^\da &=&\la_P^i \tl{P}_i P_i+\la_S \hat{S}^2+2 c_S F_\phi^\da \hat{S}~,\\
 -F_{P_i}^\da &=& \la_P^i \hat{S} \tl{P}_i + c_P^i F_\phi^\da \tl{P}_i ~,\\
 -F_{\tl{P}_i}^\da &=& \la_P^i \hat{S} {P}_i +  c_P^i F_\phi^\da {P}_i ~.
 \eeqa
Then the scalar potential is given by,
 \beqa
 V=|F_{\hat{S}}|^2+\sum\limits_{i}\(|F_P^i|^2+|F_{\tl{P}^i}|^2\)~.
 \eeqa
We can minimize the scalar potential for the scalar $S$ with the minimum of $P_i,\tl{P}_i$ satisfying $\langle P_i\rangle=\langle \tl{P}_i\rangle=0$. For simply, we will set universal $c_P^i=c_P$ and $\la_P^i=\la_P$ in our subsequent discussions. The global minimum preserves CP for $c_S<0$.
From the results in \cite{luty}, we can obtain
\beqa
\label{vev}
\langle S\rangle&=&-\f{F_\phi}{2\la_S}\(3c_S+\sqrt{c_S(c_S-4)}\),\nn\\
\langle F_S\rangle&=&\f{F_\phi}{2}\(-c_S+\sqrt{c_S(c_S-4)}\)\langle S\rangle,
\eeqa
and the effective deflection parameter $d$,
\beqa
\label{vev1}
d&=&-\f{2+\f{1}{2}X\(c_S+2-\sqrt{c_S(c_S-4)}\)}{1+X}~, \\
X&=&\f{\la_P\langle S\rangle}{c_P F_\phi}=-\f{\la_P }{2c_P\la_S}\(3c_S+\sqrt{c_S(c_S-4)}\)~.
\eeqa
Numerical result indicates that $max[3c_S+\sqrt{c_S(c_S-4)}]\approx 0.343$ with $c_S\approx -0.1213$.

With negative $c_S$ and possibly cancelation in the denominator, a relatively large deflection parameter $d$ of either sign can be realized in our scenario.
The condition $X\sim -1$ also requires that $6\la_S c_P\lesssim\la_P$. On the other hand, the condition of non-tachyonic messenger masses will also constraint the deflection parameter which will be discussed shortly.

From the scalar potential and the SUSY breaking contributions, we can obtain
the relevant mass terms for the CP-even scalar $\tl{S}$, CP-odd scalar $\tl{A}$ and the fermionic counterpart $\tilde{\psi}_S$ in the singlet superfield $\hat{S}$,
\beqa
m_{\tl{S}}^2&=& 6\la_S^2\langle S\rangle^2+4 c_S^2 |F_\phi^\da|^2+6\la_S c_S\(F_\phi^\da+F_\phi\)\langle S\rangle+ 2c_S|F_\phi|^2~,\\
m_{\tl{A}}^2&=& 2\la_S^2\langle S\rangle^2+4 c_S^2 |F_\phi^\da|^2+2\la_S c_S\(F_\phi^\da+F_\phi\)\langle S\rangle- 2c_S|F_\phi|^2~,\\
m_{\tl{\psi}_S}&=& c_S F_\phi^\da+\la_S\langle S\rangle ~.
\eeqa
 With negative $c_S$, the CP-even scalar $\tl{S}$ can be lighter than the CP-odd scalar $\tl{A}$. So in our subsequent study, we choose the CP-even scalar $\tl{S}$ as the 750 GeV diphoton resonance. It is possible that the scalar $\tl{S}$ is much lighter than $F_\phi$ while the fermionic component $\tilde{\psi}_S$ is at the order of $F_\phi$. In fact, the scalar masses $m_{\tl{S},\tl{A}}$ is determined by the explicit form of the superpotential. Certain fine-tuning may be needed to obtain such light 750 GeV $\tl{S}$ in our scenario .

The mass matrix for scalar components of messengers $(P_i~,\tl{P}^*_i)$ are typically determined by $\langle S\rangle$, $F_S$ and $F_\phi$ with
\beqa%%%%\left(P_i^*~,\tl{P}_i\right)
\left(\bea{cc} |c_P^iF_\phi+\la_P\langle S\rangle|^2 &c_P^i|F_\phi|^2-\la_P\langle F_S\rangle+2\la_P\langle S\rangle\(\la_S\langle S\rangle+c_SF_\phi\)\\ c_P^i |F_\phi|^2-\la_P\langle F_S\rangle+2\la_P\langle S\rangle\(\la_S\langle S\rangle+c_SF_\phi\) & |c_P^iF_\phi+\la_P\langle S\rangle|^2\eea\right).\nn
\eeqa%%%%\left(\bea{c}P_i\\\tl{P}_i^*\eea\right)~,
After diagonalization, we can obtain the mass eigenstates $(P_{m,i},\tl{P}_{m,i}^*)$ for the scalars
 \beqa
 m_{P_{m,i},\tl{P}_{m,i}^*}^2&=&\left|\f{}{}\la_P\langle S\rangle+c_P^i F_\phi^\da\right|^2\mp  \left|\f{}{}c_P^i|F_\phi|^2-\la_P\langle F_S\rangle+2\la_P\langle S\rangle\(\la_S\langle S\rangle+c_SF_\phi\)\right|\nn\\
 &\equiv& M^2\mp\tl{M}^2~,\nn\\
 m_{fermion}^2&=&\left|\la_P\langle S\rangle+c_P^i F_\phi^\da\right|^2\equiv M^2~,
  \eeqa
 with
\beqa
\label{threshold}
M&\equiv&\la_P\langle S\rangle+c_P^i F_\phi^\da=c_P(1+X) F_\phi~,~\nn\\
\tl{M}^2&\equiv& (d+1)M F_\phi=c_P^i|F_\phi|^2-\la_P\langle F_S\rangle+2\la_P\langle S\rangle\(\la_S\langle S\rangle+c_SF_\phi\),
\eeqa
in terms of expressions in (\ref{vev1}) and the $'-/+'$ sign corresponding to $P_{m,i}$ and $\tl{P}^*_{m,i}$, respectively.
The mass eigenstates $(P_{m,i},\tl{P}_{m,i}^*)$ are given by
\beqa
P_{m,i}=\f{1}{\sqrt{2}}(P_i+\tl{P}_i^*)~,~~\tl{P}_{m,i}^*=\f{1}{\sqrt{2}}(P_i-\tl{P}_i^*)~.
\eeqa
 In addition, the requirement that the messenger masses would not be negative\cite{luty} at the minimum requires
\beqa
\label{constraint}
(d+1)F_\phi<M.
\eeqa
which, after substituting the expressions (\ref{vev}) and (\ref{threshold}), lead to
\beqa
d+1<[c_P-\f{\la_P}{2\la_S}(3c_S+\sqrt{c_S(c_S-4)})]~.
\eeqa
We can see that the deflection parameter is bounded above to be $'c_P-1'$ in our scenario. With proper chosen $c_P$, the deflection parameter can possibly be large.

The soft SUSY broken parameters can be determined by the deflected AMSB inputs.  Assuming the effective deflection parameter is $d$, the MSSM soft SUSY broken parameters are given at the messenger scale $M$ as
\beqa
\label{gaugino}
m_{\la_i}(M)=-\f{\al_i(M)}{4\pi}F_\phi \(b_i+d N_F\).
\eeqa
with the beta function of  MSSM  $(b_1,b_2,b_3)=(-33/5,-1,~3)$ and $N_F\equiv (N_{\bf 5}+3N_{\bf 10})$. Here $N_{\bf 5}$ ($N_{\bf 10}$) denotes the number of {\bf 5}({\bf 10}) messengers, respectively.

The trilinear soft terms are given by
\beqa
\f{A_t}{F_\phi/2\pi} &=&-\f{8}{3}\al_3(M)-\f{3}{2}\al_2(M)-\f{13}{30}\al_1(M)+\f{1}{8\pi}\(6|y_t(M)|^2+{|y_b(M)|^2}\),\nn\\
\f{A_b}{F_\phi/2\pi} &=&-\f{8}{3}\al_3(M)-\f{3}{2}\al_2(M)-\f{7}{30}\al_1(M)+\f{1}{8\pi}\(|y_t(M)|^2+6{|y_b(M)|^2}+|y_{\tau}(M)|^2\),\nn \\
\f{A_\tau}{F_\phi/2\pi} &=&-\f{3}{2}\al_2(M)-\f{9}{10}\al_1(M)+\f{1}{8\pi}\(3{|y_b(M)|^2}+4|y_{\tau}(M)|^2\).
\eeqa

The sfermion masses at the messenger scale $M$ are given by
\beqa
\f{m_{\tl{F}}^2}{|F_\phi|^2}&=&\f{\al^2_3(M)}{(4\pi)^2}c_3^FG_3+\f{\al^2_2(M)}{(4\pi)^2}c_2^FG_2+\f{\al^2_1(M)}{(4\pi)^2}c_1^FG_1~,
\eeqa
in which we define
\beqa
G_i&=&\(\f{N_F}{b_i}-\f{N_F^2}{b_i^2}\) d^2+\(\f{N_F }{b_i}d+1\)^2.
\eeqa
The relevant coefficients for MSSM matter contents are given in Table \ref{tab1}.
\begin{table}[h]\caption{Coefficients for soft mass terms $(c^F_3,c^F_2,c^F_1)$}\label{tab1}
\centering
\begin{tabular}{|c|c|c|c|c|c|}
\hline
$\tl{Q}_L$     & $\tl{U}^c_L$      &$\tl{D}^c_L$         & $\tl{L}_L$         & $\tl{E}_L^c$ & $\tl{H}_d$  \\
\hline
$(8,-\f{3}{3},-\f{11}{50})$& $ (8,0,-\f{88}{25})$ &$(8,0,-\f{22}{25})$ &$ (0,-\f{3}{2},-\f{99}{50})$&$ (0,0,-\f{198}{25})$&$ (0,-\f{3}{2},-\f{99}{50})$ \\ \hline
\end{tabular}
\end{table}

The stop soft masses and Higgs masses should also include the Yukawa contributions
\beqa
\f{m_{\tl{Q}_{L,3}}^2}{|F_\phi|^2}&=&\f{m_{\tl{Q}_L}^2}{|F_\phi|^2}-\f{y_t^2}{(16\pi^2)^2}(\f{16}{3}g_3^2+3g_2^2+\f{13}{15}g_1^2-6y_t^2)~,\nn\\
\f{m_{\tl{t}_{L}^c}^2}{|F_\phi|^2}&=&\f{m_{\tl{U}^c_L}^2}{|F_\phi|^2}-2\f{y_t^2}{(16\pi^2)^2}(\f{16}{3}g_3^2+3g_2^2+\f{13}{15}g_1^2-6y_t^2)~,\nn\\
\f{m_{\tl{H}_u}^2}{|F_\phi|^2}&=&\f{m_{\tl{L}_L}^2}{|F_\phi|^2}-3\f{y_t^2}{(16\pi^2)^2}(\f{16}{3}g_3^2+3g_2^2+\f{13}{15}g_1^2-6y_t^2)~,
\eeqa

We can see that with relatively large $N_F$ and $d$, for example $d=4$ and $N_F=4$,
the gluino mass as well as the squark masses can be at order of several TeV for $F_\phi \lesssim 10$ TeV. Therefore, such a low $F_\phi$ will not conflict with the LHC constraints from the searches for the multijets with large missing energy. Moreover, since the sleptons as well as electroweakinos are always light in such scenarios, the muon $g-2$ anomaly can be solved. To demonstrate our arguments, we use the package SuSpect2 \cite{suspect2} to calculate a benchmark point for deflected AMSB without messenger-matter interactions \cite{WWYZ}. From Table \ref{tab2},
we can see that a viable soft SUSY spectrum and the 125 GeV Higgs boson mass can be obtained. Besides, such
a spectrum can satisfy the dark matter relic requirement and explain the muon $g-2$ anomaly \cite{WWYZ}.

\begin{table}[h]\caption{A benchmark point with $d>0$. All the quantities with mass dimension are in GeV.}\label{tab2}
\centering
\scriptsize
\begin{tabular}{|c|c|c|c|c|}
\hline
$N_F$     & d         & M           & $F_\phi$           & $tan\beta$ \\ \hline
10        & 1.59      & $1.09\times 10^4$    & $1.33\times 10^4$    & 15.0       \\ \hline
\hline
$m_{\tilde{H}_u}^2$   & $m_{\tilde{H}_d}^2$   & $M_1$                 & $M_2$             & $M_3$             \\ \hline
$6.98\times 10^4 $             & $1.20\times 10^5 $             & $1.82\times10^2$              & $5.48\times 10^2$          & $1.88\times 10^3$          \\ \hline
$m_{\tilde{Q}_{L}}$   & $m_{\tilde{U}_{L}}$   & $m_{\tilde{D}_L}$     & $m_{\tilde{L}_L}$ & $m_{\tilde{E}_L}$ \\ \hline
$1.30\tm 10^3$              & $1.26\tm 10^3$              & $1.26\tm 10^3$              & $3.46\tm10^2$          & $1.53\tm10^2$          \\ \hline
$m_{\tilde{Q}_{L,3}}$ & $m_{\tilde{U}_{L,3}}$ & $m_{\tilde{D}_{L,3}}$ & $A_U$             & $A_D$             \\ \hline
$1.30\tm10^3$              & $1.25\tm10^3$              & $1.26\tm10^3$              & $-6.58\tm10^2$         & $-6.50\tm10^2$         \\ \hline
$A_L$                 & $A_\tau$              & $A_t$                 & $A_b$             &                   \\ \hline
$-1.46\tm10^2$             & $-1.17\tm10^2$             &$ -2.28\tm10^2 $            &$ -5.34\tm10^2 $        &                   \\ \hline
\hline
$Br(B\rightarrow X_S\gamma)$&$Br(B_S^0\rightarrow \mu^+\mu^-)$& $g_\mu-2$           & $\Omega_\chi h^2$          & $\sigma_P^{SI}$  \\ \hline
$3.25\tm10^{-4}$                    & $3.40\tm 10^{-9}$                        & $1.82\tm10^{-9}$           & 0.117                      & $1.09\tm10^{-12}$ pb      \\ \hline
$m_{h_1}$                   & $m_{\tilde{\chi}_1^0}$          & $m_{\tilde{\tau}_1}$& $m_{\tilde{\chi}_1^{\pm}}$ & $m_{\tilde{g}}$  \\ \hline
124.4                       & 84.1                            & 100.2               & 464.5                      & 3949.4           \\ \hline
\end{tabular}
\normalsize
\end{table}

However, in ordinary deflected AMSB scenario, we should mention that Higgs mass may be lighter than 125 GeV for a very low $F_\phi$. To improve this, one can introduce additional messenger-matter interactions in the superpotential.  Such a theory  can possibly give a large $A_t$ and the 125 GeV Higgs mass with even few messenger species \cite{fei-PLB,WYZ}.

\section{750 GeV diphoton resonance in deflected AMSB scenario}
As noted previously, the 750 GeV resonance is identified as the CP-even component $\tl{S}$ of the singlet chiral superfield $\hat{S}$. The diphoton decay of $\tl{S}$ is mediated by scalar and fermion loops involving messengers. The relevant couplings between the CP-even scalar $\tl{S}$ and messengers $P,\tl{P}$ are given by
\beqa\label{lag}
-{\cal L}&\supseteq& \la_P \[ \sqrt{2}\la_P\langle S\rangle+\f{c_P}{\sqrt{2}}(F_\phi^\da+F_\phi)\]\tl{P} \tl{P}^* \tl{S}
                     +\la_P \[\sqrt{2}\la_P\langle S\rangle+\f{c_P}{\sqrt{2}}(F_\phi^\da+F_\phi)\]{P} {P}^* \tl{S}\nn\\
&& +\sqrt{2}\la_P\(\la_S\langle S\rangle+c_S F_\phi^\da\)\tl{P}P\tl{S}+\sqrt{2}\la_P\(\la_S\langle S\rangle+c_S F_\phi^\da\)(\tl{P}P)^* \tl{S}
 +\f{\la_P}{\sqrt{2}} \tl{S} \psi_{\tl{P}} \psi_P\nn\\
&=&\f{\la_P}{\sqrt{2}} \tl{S} \psi_{\tl{P}} \psi_P+\sqrt{2}\la_P M \(\tl{P} \tl{P}^* \tl{S}+ {P} {P}^* \tl{S}\)
+\sqrt{2}\la_PM_S\(\tl{P}P \tl{S}+(\tl{P}P)^* \tl{S}\),
\eeqa
with
\beqa
M&\equiv& \la_P\langle S\rangle+c_P F_\phi~, ~~~M_S\equiv\la_S\langle S\rangle+c_S F_\phi^\da~.
\eeqa
So, the relevant interactions in terms of the mass eigenstates $(P_m,\tl{P}_m^*)$ are given as
\beqa
-{\cal L}&\supseteq&\f{\la_P}{\sqrt{2}} \tl{S} \psi_{\tl{P}} \psi_P
+\f{\la_P}{\sqrt{2}} \tl{S}\[(M+M_S){P}_m^*+(M-M_S)\tl{P}_m \](P_m+\tl{P}^*_m) \nn\\
 &&+\f{\la_P}{\sqrt{2}} \tl{S}\[(M+M_S){P}_m^*-(M-M_S)\tl{P}_m \](P_m-\tl{P}^*_m)\nn\\
&=&\f{\la_P}{\sqrt{2}} \tl{S} \psi_{\tl{P}} \psi_P
+\sqrt{2}\la_P \tl{S}(M+M_S){P}_m^*P_m+\sqrt{2}\la_P \tl{S}(M-M_S)\tl{P}_m \tl{P}^*_m.
\eeqa

The diphoton decay width is given by
\beqa
 \Gamma(\tl{S}\ra \gamma\gamma)
 &=&\f{\al^2 m_S^3}{256\pi^3} N_{mess}^2\left|\sum\limits_{S=P_m,\tl{P}_m}\f{g_{\tl{S}SS}}{M_S^2}A_0\(\f{4 M_S^2}{M_{\tl{S}}^2}\)+\f{2g_{\tl{S}FF}}{M_F}A_{1/2}\( \f{4 M_F^2}{M_{\tl{S}}^2}\)\right|^2,
\eeqa
 with
 \beqa
 N_{mess}&=&\f{8}{3}N_{\bf 5}+8 N_{\bf 10}~,\nn\\
  A_{1/2}(x)&=& 2x [1+(1-x)f(x)]~,\nn\\
  A_0(x)&=&-x(1-x f(x))~,\nn\\
 f(x)&=& arcsin^2\(\f{1}{\sqrt{x}}\),~~~~x\geq 1 .
 \eeqa
where $N_{\bf 5}$ and $N_{\bf 10}$ are the numbers of $\bf{5,\bar{5}}$ and $\bf{10,\overline{10}}$ messengers, respectively.

%%%%The gluon fusion process is given by
%%%%\beqa\Gamma(gg\ra S) &=&\f{\al_S^2 m_S^3}{128\pi^3} \(\f{\la_P}{M}\)^2N_F^2\left|A_0(4 m_F^2/m_S^2)+A_{1/2}(4 m_F^2/m_S^2)\right|^2~.\eeqa
There are in total three scales in our scenario: $M,\tl{M},M_S$. The value of messenger scale $M$ appearing in Eq.(\ref{lag}) is assumed at order of 10 TeV in our scenario. The mass scale $\tl{M}$ determines the mass scale of $P_m$, which can be as low as ${\cal O}(TeV)$ while the upper bound for $\tl{P}_m$ is approximately $\sqrt{2}M$. We consider the following two cases in our numerical results:
\bit

\item  [{\bf A}.]     The masses of messenger scalars $P_m,\tl{P}_m$ are set to be $m^2_{P_m}=(2~{\rm TeV})^2$ and $m_{\tl{P}_m}^2\approx(\sqrt{2}M)^2$.
            The mass scale of $M_S$ is typically at the same order of $M$ and we set $M_S=0.5 M$ for simplicity.
            The corresponding Yukawa coupling $g_{\tl{S}FF}$ and trilinear coupling $g_{\tl{S}SS}$ are taken as
 \beqa
  g_{\tl{S}FF}=\f{\la_P}{\sqrt{2}}~,~g_{\tl{S}P_mP_m^*}=\sqrt{2}\la_P( 1.5 M)~,~g_{\tl{S}\tl{P}_m\tl{P}^*_m}=\sqrt{2}\la_P (0.5 M).
\eeqa

\item  [{\bf B}.]   The messengers (fermions and scalars) are set to have a common mass $M_P\equiv M$ ($\tl{M}\ll M$) and also $M_S\ll M$. The relevant couplings are
\beqa
  g_{\tl{S}FF}=\f{\la_P}{\sqrt{2}}~,~g_{\tl{S}SS}=\sqrt{2}\la_P M~.
\eeqa
\eit

%%fig.1 %%%%%%%%%%%%%%%%%%%%%%%
\begin{figure}[ht]
\centering
\includegraphics[width=4 in]{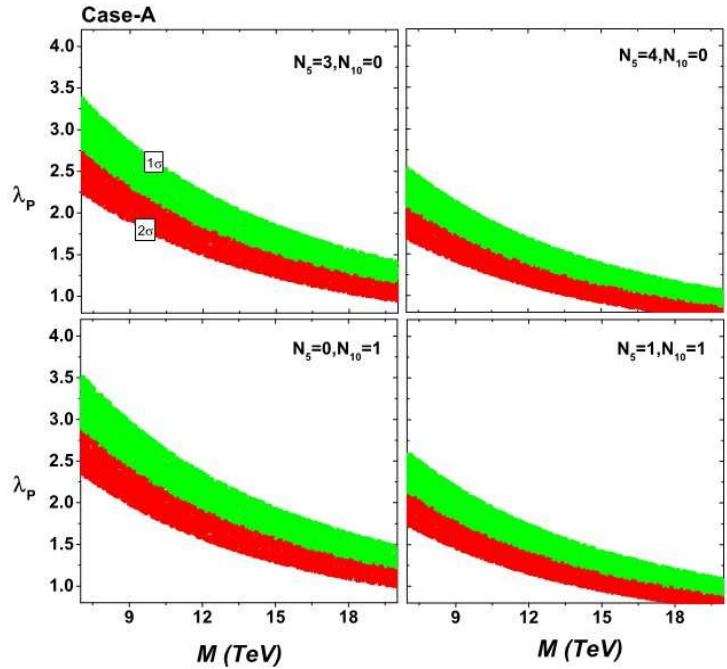}
\caption{The scatter plot on the plane of the messenger scale $M$ and $\la_P$ under different choices of $5,\bar{5}$($10,\overline{10}$ )
messengers for case-A. The lightest scalar messenger mass is assumed as $m^2_{P_m}=(2~{\rm TeV})^2$. The green and red bullets correspond to $1\sigma$ and $2\sigma$ range of Eq.(\ref{excess}), respectively.}
\label{result1}
\end{figure}
%%%%%%%%%%%%%%%%%%%%%%%%%
We scan the Yukawa coupling $\lambda_{P}$, the number of messengers $N_{mess}$ and
the messenger scale $M$ within the following ranges,
\begin{eqnarray}
0 \leq \lambda_P \leq 4\pi, \quad 7 ~{\rm TeV} \leq M \leq 20 ~{\rm TeV}.
\end{eqnarray}
In our scan, we require our samples to explain the diphoton excess in $2\sigma$ range of Eq.(\ref{excess}) and satisfy the following constraints:
\begin{itemize}

\item [(1)] The CMS search for a dijet resonance~\cite{jj} at $\sqrt{s}=8$ TeV
with ${\cal L} =18.8~fb^{-1}$ gives a $95\%$ C.L. upper limit on the production
of the RS graviton decaying to $gg$
\be
	\sigma(p p \to X)_{\rm{8\,TeV}} \times Br(X\to g g) < 1.8~\text{pb}~.
	\label{ggbound}
\ee
\item [(2)] The ATLAS~\cite{vv-atlas} and CMS~\cite{vv-cms} searches for a scalar resonance
decaying to $VV(V=W,Z)$ at $\sqrt{s}=8$ TeV with the full data set, combining all relevant
$Z$ and $W$ decay channels, give a $95\%$ CL upper limit on the production of the scalar
decaying to $VV$
\begin{align}
&\sigma(p p \to S)_{\rm{8\,TeV}} \times \mathcal{B}(S\to Z Z) < 22\,{\rm fb}\,_{\rm(ATLAS)} \,,~ 27\,{\rm fb}\,_{\rm(CMS)}\,,\label{ZZbound}\\
&\sigma(p p \to S)_{\rm{8\,TeV}} \times \mathcal{B}(S\to W W) < 38\,{\rm fb}\,_{\rm(ATLAS)} \,,~ 220\,{\rm fb}\,_{\rm(CMS)}\,.\label{WWbound}
\end{align}

\item [(3)] The ATLAS~\cite{aa1} and CMS~\cite{aa2} searches for a resonance decaying to $\gamma\gamma$
at $\sqrt{s}=8$ TeV give a $95\%$ CL upper limit
\begin{align}
&\sigma(p p \to X)_{\rm{8\,TeV}} \times Br(X\to \gamma\gamma) < 2.2\,{\rm fb}\,_{\rm(ATLAS)} \,,~ 1.3\,{\rm fb}\,_{\rm(CMS)}\, .\label{aabound}
\end{align}
\end{itemize}
We calculate the production cross section $gg \to S$ at the 13 TeV LHC by using the
package \textsf{HIGLU} \cite{higlu} with CTEQ6.6M PDFs \cite{cteq6}.
The renormalization and factorization scales are set as $\mu_R=\mu_F=m_S/2$. We also include a
$K_{gg}$ factor  to account for the higher order QCD corrections in the calculation of the decay width of $S \to gg$.

In  Fig.\ref{result1}, we present scatter plot on the plane of the messenger scale $M$ and $\la_P$ under different choices of $5,\bar{5}$($10,\overline{10}$) messengers for case-A. The lightest scalar messenger mass is assumed as $m^2_{P_m}=(2~{\rm TeV})^2$. The green and red bullets correspond to $1\sigma$ and $2\sigma$ range of Eq.(\ref{excess}), respectively. All samples are required to satisfy the LHC constraints (1)-(3). For case-A, the dominant contributions for diphoton decay come from the light scalar $P_m$ loops. Due to the enhanced scalar couplings, the Yukawa coupling $\la_P$ can be of ${\cal O}(1)$ for messenger scales $M\sim {\cal O}(10 TeV)$, which requires certain fine-tuning between $M$ and $\tl{M}$ to obtain the light $P_m$. From Fig.\ref{result1}, we can also see that the Yukawa coupling $\la_P$ become smaller when the generation of messenger field increases for the same messenger scale. We find that the most stringent bound comes from the diphoton resonance measurement at the 8 TeV LHC. This produces an upper limit $\sim 5$ fb on the production rate of $gg \to S \to \gamma\gamma$ at the 13 TeV LHC.

  On the other hand, large trilinear coupling in case-A with light messenger scalars at the IR region could cause the formation of bound states for scalar messengers\cite{strong,strong1}. In fact, an attractive force between the messengers can cause such formation of bound states by exchanging the intermediate scalar $S$ particle as long as the large trilinear coupling exceed some critical value $\la_c$. Similar phenomenon can happen in the MSSM for strong trilinear interaction $A_t \tl{Q}_L H_u \tl{t}_R$\cite{strong}.  In our scenario, the light scalar messengers can form color-singlet tightly bound states with the lowest lying binding energy controlled approximately by $(\la_P M)^2/\sqrt{m_{P_m}m_{\tl{P}_m}}$. Such bound state can mix with the scalar $S$ and lead to direct coupling of gluons to the mixed mass eigenstates. It would be very interesting to explore the relevant phenomenology with non-perturbative techniques.
%%fig.2 %%%%%%%%%%%%%%%%%%%%%%%
\begin{figure}[ht]
\centering
\includegraphics[width=4 in]{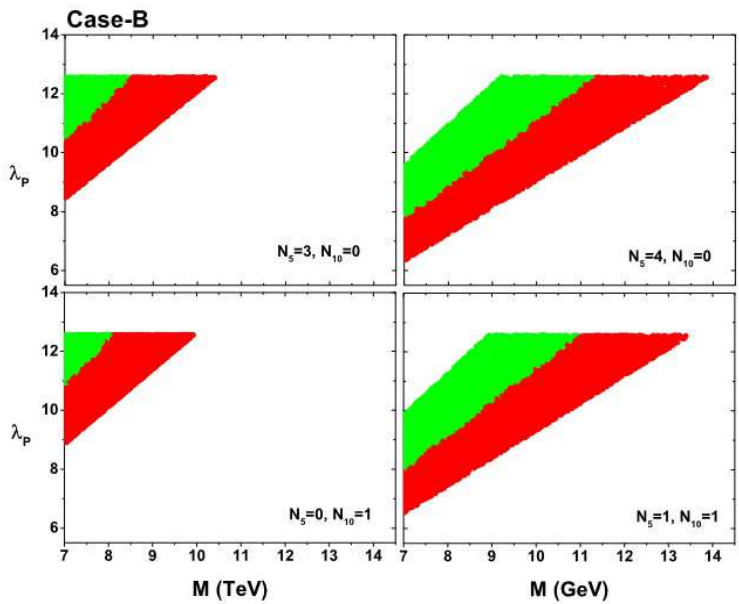}
\caption{Same as Fig.\ref{result1} But the messengers (fermions and scalars) are assumed to have a common mass $M_P\equiv M$ and $M_S\ll M$.}
\label{result2}
\end{figure}
%%%%%%%%%%%%%%%%%%%%%%%%%

Same as Fig.\ref{result1}, we present scatter plot on the plane of the messenger scale $M$ and $\la_P$ under different choices of $5,\bar{5}$($10,\overline{10}$) messengers for case-B. Such a scenario corresponds to the SUSY limits. Therefore, the large Yuakawa coupling $\lambda_{P}>6$ and large $N_F$ are required to enhance the cross section of $gg \to S \to \gamma\gamma$ to satisfy the $2\sigma$ range of Eq.(\ref{excess}). Different from case-A, the Yukawa coupling $\lambda_P$ can be small only if the number of messenger species is large. In this case, the gauge coupling will become strong at the unification scale.

\section{Conclusion}
 We proposed to interpret the 750 GeV diphoton excess in deflected anomaly mediation SUSY breaking scenarios, which can naturally predict the coupling between a singlet field and the vector-like messengers. The most general form with possibly holomorphic Kahler potential and messenger-matter interactions were discussed. It is crucial that the gravitino scale $F_\phi$, which determine the whole spectrum, can be at order or less than
10 TeV without contradicting with the LHC constraints when the messenger species number $N_F$ as well as
the deflection parameter $d$ are moderately large. The CP-even scalar component of the singlet, whose mass is
model-dependent, can be light and serve as the 750 GeV resonance while its fermionic component can be heavy.
The messenger fields can induce the large loop decay process $S \to \gamma\gamma$.
 Our results show that such a scenario can successfully accommodate the 125 GeV Higgs boson, 750 GeV diphoton excess and
the muon $g-2$ anomaly without conflicting with the LHC constraints.

We should comment on the possibility to interpret the diphoton excess in the GMSB scenario.
One can in principle introduce an additional light singlet field (other than the hidden sector singlet $X$) that couples to messenger fields in
the GMSB scenario. However, the gravitino mass which set the $F_X$ scale is stringently constrained.
A light gravitino can be problematic in cosmology because there is a severe upper bound on the reheating
temperature from the requirement that the gravitinos do not overclose the universe. As pointed out
in \cite{gravitinoGMSB},  the gravitino with mass below electroweak scale and $m_{3/2}> {\cal O}(10)$eV
can cause such cosmological problems. Low-scale SUSY breaking with a gravitino mass as light as $1-16$ eV
is allowed, which, however, will in general encounter the constraint from vacuum instability and most cases
are already excluded by SUSY searches at the LHC. Even for eV scale gravitino, as $F_X$ and $\langle X\rangle$
will determine the whole soft SUSY parameters, the constraints on $F_X$ and LHC discoveries will set the
scale $\langle X\rangle$ of order 100 TeV. Such heavy messengers will in general decouple and play no roles
in explaining the diphoton excess.

\section{Acknowledgments}
We are very grateful to the referee for helpful discussions. This work is supported by the National Natural Science Foundation of China (NNSFC)
under grants Nos. 11105124,11105125,\\
11275057, 11305049, 11375001, 11405047, 11135003,
11275245, by the Open Project Program of State Key Laboratory of Theoretical Physics, Institute of Theoretical Physics, Chinese Academy of Sciences
(No.Y5KF121CJ1), by the Innovation Talent project of Henan Province under grant number
15HASTIT017, by the Joint Funds of the National Natural Science Foundation of China
(U1404113), by the Outstanding Young Talent Research Fund of Zhengzhou University(1421317054,1421317053), and by the CAS
Center for Excellence in Particle Physics (CCEPP).

\end{document}